# CHANNELING AND VOLUME REFLECTION BASED CRYSTAL COLLIMATION OF TEVATRON CIRCULATING BEAM HALO (T-980)*

V. Shiltsev[#], G.Annala, R.Carrigan, A.Drozhdin, T.Johnson, A.Legan, N. Mokhov, R.Reilly, D.Still, R.Tesarek, J.Zagel, FNAL, Batavia, IL 60510, U.S.A.
S.Peggs, BNL, Upton, NY 11973, U.S.A.
R.Assmann, V.Previtali, W.Scandale, CERN, Geneva, CH-1211, Switzerland
Y.Chesnokov, I.Yazynin, IHEP, Protvino, Moscow Region, RU-142284, Russia
V.Guidi, INFN-Ferrara, Italy
Y.Ivanov, PNPI, Gatchina, Leningrad Region, RU-188300, Russia

*Abstract*

The T980 crystal collimation experiment is underway at the Tevatron to determine if this technique could increase 980 GeV beam-halo collimation efficiency at high-energy hadron colliders such as the Tevatron and the LHC. T980 also studies various crystal types and parameters. The setup has been substantially enhanced during the Summer 2009 shutdown by installing a new O-shaped crystal in the horizontal goniometer, as well as adding a vertical goniometer with two alternating crystals (O-shaped and multi-strip) and additional beam diagnostics. First measurements with the new system are quite encouraging, with channeled and volume-reflected beams observed on the secondary collimators as predicted. Investigation of crystal collimation efficiencies with crystals in volume reflection and channeling modes are described in comparison with an amorphous primary collimator. Results on the system performance are presented for the end-of-store studies and for entire collider stores. The first investigation of colliding beam collimation simultaneously using crystals in both the vertical and horizontal plane has been made in the regime with horizontally channeled and vertically volume-reflected beams. Planning is underway for significant hardware improvements during the FY10 summer shutdown and for dedicated studies during the final year of Tevatron operation and also for a "post-collider beam physics running" period.

## INTRODUCTION

An efficient beam collimation system is mandatory at any collider and high-power accelerator. A common approach is a two-stage collimation system in which a primary collimator is used to increase the betatron oscillation amplitudes of the halo particles, thereby increasing their impact parameters on secondary collimators. A bent crystal can coherently direct channeled halo particles deeper into a nearby secondary absorber. This approach has the potential of reducing of beam losses in critical locations and radiation loads to the downstream superconducting magnets. Extended literature on earlier theoretical and experimental works on the use of a bent crystal for beam halo collimation as well as first results of the T980 experiment can be found in Ref.[1].

Important insights into the physics of crystal deflected beams has been made at CERN at the H4/H8 beamlines [2,3]. Using 400 GeV/c extracted beams, these experiments experimentally confirmed that there are several processes which can take place the during passage of protons through the crystals: a) amorphous scattering of the primary beam b) channeling; c) dechanneling due to scattering in the bulk of the crystal; d) "volume reflection" off the bent planes; and e) "volume capture" of initaily unchanneled particles into the channeling regime after scattering inside the crystal. The process of "volume reflection" (VR) is very attractive for the purpose of the halo beam collimation as it has significantly larger angular acceptance of the order of hundreds of microradians compared with the acceptance of channeling (CH) which is of the order of the critical angle $+\text{-}\theta_c$~5-10 microradians at the TeV energies. VR also has very high efficiency. The drawback of the volume reflection regime is that the deflection angle is small, of the order of $(1.5\text{-}2)\times\theta_c$. However, this can be overcome by using a sequence of several precisely aligned bent crystals, so the total deflection angle is proportionally larger [3].

In the T980 experiment both single crystals (for vertical and horizontal deflection) and multi-strip crystal assemblies (for vertical multiple VR) are used. Another significant difference between the CERN H4/H8 studies is that in the Tevatron the beams are circulating. That makes a pivotal difference because of smaller initial "impact parameters" and the possibility of interplay of different effects. In an accelerator such as the Tevatron several phenomena determine the impact parameter (the depth of the particle penetration at the first interaction with the crystal).These include four diffusion and two orbit processes.The diffusion processes are scattering on vacuum molecules and transverse noise (~4 nm/√turn); RF noise (~12 nm/√turn (hor) and ~1 nm/√turn (vert));

___________________________________________
*Work supported by Fermi Research Alliance, LLC, under contract No. DE-AC02-07CH11359 with the U.S. Department of Energy through the US LHC Accelerator Research Program (LARP
[#]shiltsev@fnal.gov

beam-beam or other nonlinear beam diffusion (~10-40 nm/√turn); and TEL-driven particles in the abort gap (~7 µm/turn). For interaction with amorphous targets, these diffusion rates are ~200 µm/√turn for a 5 mm amorphous Si, and about ~1200 µm/√turn for a 5 mm W primary target. The two orbit processes are : transverse orbit oscillations with amplitude of ~20 µm and frequencies of ~15 Hz (that's some 3000 revolutions in the Tevatron) and synchrotron motion of particles near the boundary of the RF bucket with amplitudes of ~1 mm (hor) and ~70 µm (vert) at ~35 Hz (1300 turns). The later is true for non-zero dispersion locations , e.g. about 2m horizontal and some 18 cm vertical at the E03 location of the Tevatron crystals. The resulting impact parameters are estimated to be of the order of 0.2-1 µm for transverse halo particles and ~10-30 µm for the particles in the abort gaps which have leaked out of the RF buckets. All that makes the properties of the surface of the crystal (rather than the bulk of the crystal) pivotal for collimation (contrary to the extracted beam studies) – e.g. it's roughness or the miscut angle [1]. It is not easy to model these processes in simulations and, hence, the T-980 experimental studies are of great importance.

## EXPERIMENTAL STUDIES

A schematic of the T-980 experimental layout and a detailed description can be found in Ref.[1]. Since the time of that report, the setup has been substantially enhanced during the Summer 2009 shutdown by installing a new O-shaped crystal in the horizontal goniometer, adding a vertical goniometer with two alternating crystals (O-shaped and multi-strip) and by additional beam loss diagnostics at location of collimators at E03 and F17 sectors of the machine. Most of the studies reported below have been performed with a horizontal O-shaped crystal with a 120µrad miscut angle and a bending channeling angle of 360 µrad (inward toward the center of the Tevatron ring) and a vertical 8-strip crystal with deflection downward by 200 µrad in the channeling regime and deflection upward by about 60 µrad in VR regime. The vertical and horizontal goniometers are installed near each other at the E0 location.

The beam studies were usually performed at the end of Tevatron stores and only with protons. They were of two types: angular scans of the crystal and transverse collimator position scans of the extracted beams. In addition to standard log-amp Tevatron BLMs which integrate over 1/16 sec [4], an enhanced beam diagnostics system has been installed that made it possible to distinguish the effects of the bunched beam and beam in the abort gap in several locations - at the collimators E03V and H (about 24 m downstream of the crystals), at F17V and H some km downstream where deflected beam can be intercepted, at CDF and D0 detectors (further 2 and 4 km downstream, respectively).

Angular scans to set the crystal in the channeling configuration are performed by measuring losses at E03 while the orientation of the crystal is changed (Fig. 1).

The E03 collimator distance from the beam is set to be slightly larger than that of the crystal. In principle, one expects to see a loss peak on the E03 collimator when the halo particles are aligned with the crystalline planes at the entrance face of the crystal. Full angular acceptance for channeling is predicted to be about $1.5\times \theta_c$ =10 µrad. In reality, the acceptance was always large than that, eg, 7.4 µrad rms or some 25-30 µrad full width for the scan presented in Fig.1.

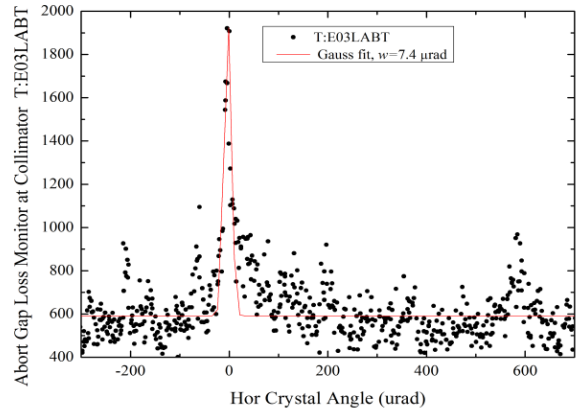

Figure 1: Results of horizontal crystal angular scan.

Under similar conditions of limited aperture, i.e. with E03V and F172V collimators set close to the beam, an angular scan has been performed with the 8-strip vertical crystal (Fig.2). As qualitatively expected from the STRUT simulations, the loss rate measured by the LE033 BLM increases when upward deflected VR beam hits the E03V collimator (for angles between 0 and 200 µrad) and decreases when the CH beam is deflected downward (away from the collimator) at angles from -100 to 0 µrad.

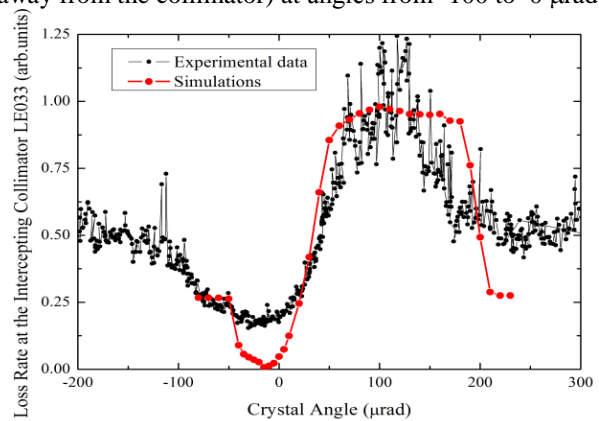

Figure 2: Results of angular scan with 8-strip vertical crystal. Red line is for computer simulations results.

To measure the deflection of the channeled (or VR) particles once the crystal angle is set to the channeling (or VR) peak the position of an appropriate collimator is slowly scanned, starting from a completely retracted position and moving toward the beam edge. An example of such scan is shown in Fig.3 for horizontally deflected CH beam. at the E03H collimator. The curves show total losses (red dots) as well as for synchronized to the abort gaps (black dots) . There are three distinct regions: a) region of negligible losses, where the collimator does not intercept any beam; b) a steep increase in the losses, where the collimator intercepts the channeled beam; c) a region where the losses increase slowly: the collimator is additionally intercepting dechanneled and amorphous scattered particles.

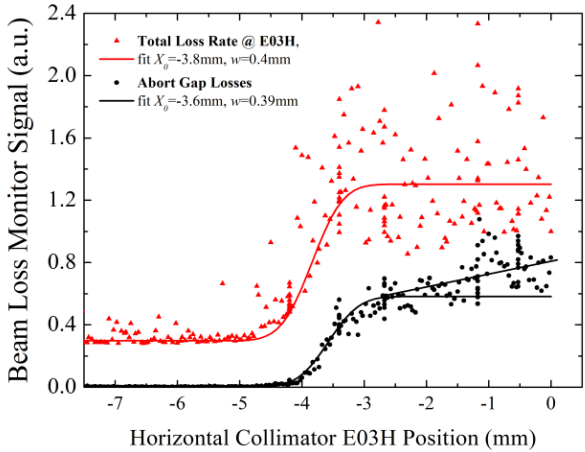

Figure 3: Collimator scan with crystal set at the channeling angle. Solid lines are for "erfc" fit of the data.

Both abort gap and total loss signals show a small deflection angle of (3.6-3.8)mm/24 m=150-160 μrad instead of expected 360 μrad; and large angular spread in the CH beam of about of 0.4 mm/24 m=17 μrad rms that is quite larger than the CH acceptance of 13.4 μrad.

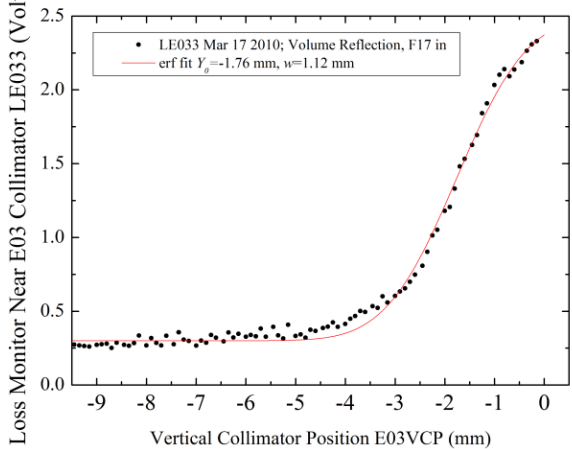

Figure 4: Collimator scan with 8-strip vertical crystal set at the VR angle. Solid line is for "erfc" fit of the data.

A similar scan of the VR beam made with the E03 vertical collimator (Fig.4) shows the beam at 1.76mm/28m=63 μrad, i.e. approximately where it is supposed to be; and about 40 μrad wide (rms).

Finally, we have investigated the two plane collimation by setting the vertical multi-strip crystal at the VR angle and the horizontal O-shaped crystal at the CH angle (W targets retracted, secondary collimators in), and observed that at least horizontal collimation has efficiently reduced abort gap losses ringwide (see yellow line in Fig.5).

## SUMMARY AND PLANS

In the recent 2009-2010 T-980 studies we have routinely employed crystal collimation during many collider stores. In 2009 the old O-shaped crystal in the horizontal goniometer was replaced with the new 0.36-mrad O-shaped one with negative 0.12-mrad miscut angle (built by IHEP) and a new vertical push-pull goniometer was installed 4-m upstream, housing two crystals - the 8-strip (IHEP) and the old O-shaped ones, so that we now have crystals for both V and H planes. Additional beam instrumentation was added. Fast/automatic insertion of crystals has been implemented. We have successfully tested a vertical multi-strip crystal system and observed both multiple-VR beam at the E03 collimator and channeled beam at the F17 collimator. A reduction of ring wide losses was reproducibly observed along with local loss effects on the collimator due to crystal channeling and VR. We have conducted the first ever study of two plane crystal collimation. There are quantitative discrepancies between simulations and observations which will be addressed in the future studies.

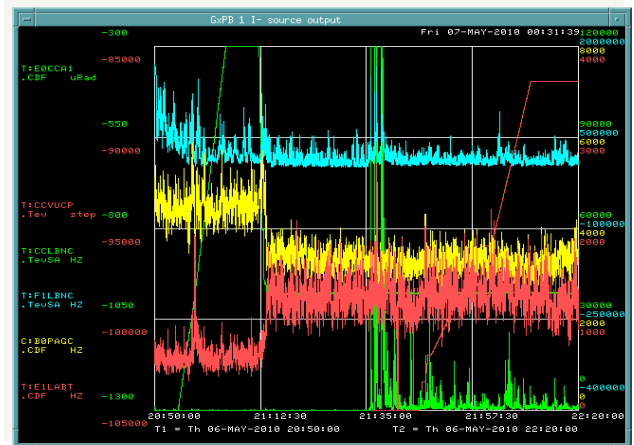

Figure 5: Beam study with 2-plane crystal collimation.

In the summer 2010 shutdown the old O-shaped crystal in the vertical goniometer will be replaced with a new Quasi-Mosaic crystal (made at PNPI), and an 8-strip IHEP crystal will be replaced with an advanced 16-strip crystal (made by INFN, Ferrara), keeping a possibility to alternate them remotely. High-resolution pixel telescopes will be installed in front of E03 and F17 collimators to measure channeled and VR beam profiles at those locations with a 5 μm resolution. A broad experimental program with this enhanced system is planned in 2010-11 period for thorough studies of two-plane crystal collimation efficiency, demonstration of improved reproducible beam loss localization in collimation regions, and reduction of beam losses around the ring, specifically near the CDF and D0 detector regions.